# Composition Dependence of Structural Parameters and Properties of Gallium Ferrite

*Somdutta Mukherjee[1], Vishal Ranjan[2], Rajeev Gupta[1,3] and Ashish Garg[2*]*

[1]Department of Physics

[2]Department of Materials Science and Engineering

[3]Materials Science Programme

Indian Institute of Technology Kanpur, Kanpur 208016, India

**Abstract:**

We show the effect of composition on structural and magnetic characteristics of pure phase polycrystalline $Ga_{2-x}Fe_xO_3$ (GFO) for compositions between $0.8 \leq x \leq 1.3$. X-ray analysis reveals that lattice parameters of GFO exhibit a linear dependence on Fe content in single phase region indicating manifestation of Vegard's law. Increasing Fe content of the samples also leads to stretching of bonds as indicated by the Raman peak shifts. Further, low temperature magnetic measurements show that the coercivity of the samples is maximum for Ga:Fe ratio of 1:1 driven by a competition between decreasing crystallite size and increasing magnetic anisotropy.

---

[*] Corresponding author; Email: ashishg@iitk.ac.in

Recent years have witnessed a large number of investigations on magnetoelectric and multiferroic materials due to their potential in novel device applications such as multi-state memory elements and spin field effect transistors.[1-6] Gallium ferrite (GaFeO$_3$ or GFO) is a room temperature piezoelectric and a ferrimagnet with transition temperature tunable to room temperature or higher.[7] The material also shows significant magnetoelectric coupling.[8,9] Interestingly, site exchanges between Ga and Fe atoms which have size difference of ~4% lead to cationic site disorder resulting in ferrimagnetism in GFO.[9,10] This characteristic also implies that tuning of Ga:Fe ratio in the material could play an important role in tailoring its properties.[9]

Typically, materials like GFO which demonstrate large magnetoelectric coupling also show substantial magnetostructural coupling making these attractive for a variety of applications. In this context, it would be important to revisit the effect of composition tuning on the structure and phase evolution as well as magnetic behavior of GFO. GFO has a non-perovskite and a non-centrosymmetric orthorhombic structure with space group $Pc2_1n$ and its room temperature lattice parameters are a ~ 8.75 Å, b ~ 9.40 Å, c ~ 5.08 Å.[11] The unit-cell consisting of eight formula units has two nonequivalent Ga sites and Fe sites, marked as Ga1, Ga2 and Fe1, Fe 2, respectively. Here, Ga1 is tetrahedrally coordinated while Ga2, Fe1 and Fe2 are octahedrally coordinated by oxygen. FeO$_6$ octahedra are significantly distorted and the shift of Fe1 and Fe2 cations from the center of octahedra along *b*-axis renders GFO a polar material.[9] As far as magnetism in GFO is concerned, Mossbauer spectroscopy[10] and subsequent neutron diffraction studies[9] confirmed GFO as a ferrimagnet with cation site disorder as source of magnetism. In this manuscript, we show that structural parameters of GFO follow Vegard's law very well. We also find that samples exhibit a rather broad range of composition where single phase GFO is stable. Further the low temperature magnetic measurements show that while remanent magnetization has linear compositional dependence, coercive field attains a maximum for samples with $x = 1.0$.

Polycrystalline $Ga_{2-x}Fe_xO_3$ ($0.7 \leq x \leq 1.4$) ceramic samples were prepared using conventional solid-state-reaction method by mixing high purity $\beta$-$Ga_2O_3$ and $\alpha$-$Fe_2O_3$ powders in relevant proportions and subsequently calcining the mixtures between 1300-1400°C in air. Powder X-ray diffraction (XRD) of the samples was conducted on Philips X'Pert Pro MRD X-Ray diffractometer using Cu $K\alpha$ radiation. Raman spectra were acquired in backscattering geometry using a JY-Horiba micro Raman system equipped with a $Ar^+$ Laser and Peltier cooled CCD detector. Temperature dependent magnetization measurements were carried out using a vibrating sample magnetometer in the temperature range 115-400 K with a magnetic field of 500 Oe. Low temperature magnetic hysteresis measurements were performed at 115 K.

Room temperature XRD patterns for all the compositions of $Ga_{2-x}Fe_xO_3$ ($0.7 \leq x \leq 1.4$) are shown in Fig. 1(a). The compositions used were $x = 0.7, 0.8, 0.9, 1.0, 1.1, 1.2, 1.3$ and 1.4 which correspond to 35, 40, 45, 50, 55, 60, 65 and 70 mol% $Fe_2O_3$ in $(Ga_2O_3)_{(2-x)/2}(Fe_2O_3)_{x/2}$. First, the peaks in the XRD spectra were matched with $GaFeO_3$ (ICDD card no: 761005) and it was found that the samples with compositions from $x = 0.8$ to 1.3 show single phase of GFO, while samples with two extreme compositions $x = 0.7$ and 1.4 showed presence of unreacted $\beta$-$Ga_2O_3$ (ICDD Card No. 871901) and $\alpha$-$Fe_2O_3$ (ICDD Card No. 890599) respectively, in addition to GFO. This suggests that single phase GFO exists between $0.8 \leq x \leq 1.3$ and this range is also used for further analysis of the samples. Further we carried out Rietveld analysis of the XRD data using FullProf 2000 package[12] and considering orthorhombic $Pc2_1n$ symmetry for $GaFeO_3$ and monoclinic $C2/m$ symmetry for $\beta$-$Ga_2O_3$ and rhombohedral $R\bar{3}c$ symmetry for $\alpha$-$Fe_2O_3$. The refinement results are plotted as solid lines along with the raw data in Fig 1(a). The refinement yielded a goodness of fit parameter varying between 1 to 3 for the entire composition range. Further phase analysis of the refined data suggests that the sample with $x = 0.7$ consists of ~80 mol% of the orthorhombic GFO and ~20 mol% as $\beta$-$Ga_2O_3$. On the other hand, sample with $x = 1.4$ shows presence of 74 mol% GFO and ~26 mol% of $\alpha$-$Fe_2O_3$. These two compositions ($x \geq 0.8$ and $\leq 1.3$) mark the phase boundaries for the existence of single phase GFO. These results are in excellent agreement with the previously reported data on the phase stability of $GaFeO_3$.[13] The results also suggest that GFO is not a line compound as it can accommodate cation non-stoichiometry from $x = 0.8$ to 1.3 without formation of any secondary phases or any un-reacted ingredients. From the perspective of Gibbs phase rule, it means that at room temperature, the variation in composition does always yield single phase GFO within these compositional limits.

Further, if we observe Fig. 1(b), we see that the peak position of the highest intensity peak (2θ ~ 33.03°) shifts towards lower 2θ with increasing Fe content for single phase GFO indicating an increase in the lattice parameters with increasing Fe content. Similar trend was observed for other peaks too. We then determined the lattice constants of GFO as a function of composition from the refined XRD data which are plotted in Fig 2. Here we see that the lattice constants ($a$, $b$, and $c$) as well as the cell volume increase with increasing linearly Fe content. The data shown is only for the composition $0.8 \leq x \leq 1.3$ corresponding to pure GFO phase. Such a linear variation of lattice constants with composition in a binary system can be quantitatively expressed by Vegard's law or a simple rule of mixtures. However, since any further change in composition beyond $0.8 \leq x \leq 1.3$ leads to unreacted starting powders, the extrapolation of line does not fall anywhere close to lattice constants of $\beta$-$Ga_2O_3$ and $\alpha$-$Fe_2O_3$. Also, $\beta$-$Ga_2O_3$ and $\alpha$-$Fe_2O_3$ have monoclinic and hexagonal structures respectively, significantly different than GFO's structure. For a linear extrapolation until the end members, one needs to use the lattice parameters of isostructural forms of reactants *i.e.* $\kappa$-$Ga_2O_3$ and $\varepsilon$-

Fe$_2$O$_3$ as end members (for $x = 0$ and $x = 2$, respectively). The lattice parameters of these two compounds were obtained from references [14-16]. The lattice parameters of κ- Ga$_2$O$_3$ were corrected to room temperature values by using the thermal expansion coefficients of GFO.[17] A comparison of these lattice constants with those obtained from the extrapolation of line in Fig. 2 to $x = 0.0$ and $2.0$ reveals an excellent correspondence and is a clear manifestation of Vegard's law. Hence, based on these observations, a relation between the lattice parameters of κ- Ga$_2$O$_3$, Ga$_{2-x}$Fe$_x$O$_3$ ($0.8 \leq x \leq 1.3$) and ε-Fe$_2$O$_3$ can be expressed as

$$a_{i(i=1,2,3),Ga_{2-x}Fe_xO_3} = \frac{(2-x)}{2} a_{i(i=1,2,3),\kappa-Ga_2O_3} + \frac{x}{2} a_{i(i=1,2,3),\varepsilon-Fe_2O_3}$$

With the exception of a few systems such as YMn$_{1-x}$Ga$_x$O$_3$[18], well-studied magnetoelectric materials have not been reported to obey Vegard's law. As far as perovskite structured oxides are concerned, while solid solutions with structures close to cubic structure[19] e.g. BaTiO$_3$-CaTiO$_3$, BaTiO$_3$-BaZrO$_3$, and SrTiO$_3$-BaZrO$_3$ follow Vegard's law very well, there have been many reports of deviations in some other perovskite oxides[20,21]. In this context, it is indeed remarkable to observe Vegard's law in the single phase region of Ga$_2$O$_3$-Fe$_2$O$_3$ system. Unlike simple perovskite oxides, GFO is a crystallographically complex oxide having orthorhombic structure with 40 atoms per unit-cell. With significantly different lattice dimensions, the prediction of lattice constants as a function of composition allows one to tailor the material for epitaxial films for devices. The increase in the unit cell volume is about ~2.5%, which is expected as the ionic radius of Fe$^{3+}$ (~ 0.645 Å) is 4% larger than that of Ga$^{3+}$ (~0.62Å). This large change in the ionic radius also affects the polyhedral distortion in the unit cell. This distortion can be quantified using Baur's[22] distortion index,

$$\Delta = \frac{1}{n} \sum_{i=1}^{n} \frac{(l_i - l_{avg.})}{l_{avg.}}$$ where $l_i$ is distance from the central atom to the $i^{th}$ coordinating atom, $n$ is the number of bonds and $l_{avg.}$ is average bond length. The calculations suggest that while $\Delta$ is relatively large for Fe1 and Fe2 octahedra, change is maximum for Ga2 octahedra upon

increasing the Fe content. In contrast, the distortion index of Ga1 tetrahedra is about an order of magnitude smaller and remains almost unchanged with varying Fe content. Latter also appears to substantiate the fact that Fe substitution predominantly occurs on octahedral Ga2 sites.[9,23]

To further probe any subtle changes in the structure as a consequence of changes in the composition, we obtained Raman spectra of our samples. As shown in figure 3, unpolarized Raman spectra of polycrystalline specimens yielded 24 Raman active modes in the spectral range of 80-870 cm$^{-1}$ at room temperature. Although the observed number is much smaller than the number predicted by the group theory, it is consistent with earlier reports.[17] From a cursory examination of these spectra, we observe that polycrystalline nature of our samples leads to suppression of some of the modes due to broadening. Further, a comparison of these spectra for various compositions shows that number of observed Raman modes remains same upon increasing the Fe content of the samples suggesting absence of any structural change upon changing the composition. However, a closer examination reveals that the most of the modes soften with increasing Fe content, as clearly observed in the inset of Fig. 3. The inset show that the position of modes between 600-850 cm$^{-1}$ moves toward lower frequency side as Fe content increases from $x$ = 0.8 to 1.3.

Change in the mode frequency with increasing Fe content of the samples can be correlated to the changes in the bond strength, cation size and mass of cations. Ionic bond strength of Ga-O and Fe-O bonds are expected to remain unchanged as electronegativities ($\chi$) of Ga and Fe are quite similar (Bond strength $\sigma \propto |\chi_c-\chi_a|$, $c$: cation, $a$: anion). However, the increase in the cation size by ~4 % upon increasing Fe content (Fe$^{3+}$ ~ 0.645 Å and Ga$^{3+}$ ~ 0.62Å) results in lattice expansion and hence, an increase in bond length as well as distortion index as also shown by the XRD results. This increase in the bond length ($l$) implies a decrease in the phonon frequency $\omega$ ($\omega \propto l^{-\frac{1}{2}}$) as observed in our experimental data. The contribution of the differences in the atomic weights of Fe (56) and Ga (70) has a negligible effect on the phonon frequencies in the region 230-750 cm$^{-1}$ as these involve motion of oxygen polyhedra. On the other hand, the low frequency modes show no dispersion with increasing Fe content. This is possibly due to two competing mechanisms: increase in lattice parameter leading to mode softening while a decrease in effective atomic mass (as Fe content is increased) would imply mode hardening.

Following the structural characterization of GFO, we carried out magnetic measurements to investigate the magnetic parameters and transition temperature vis-à-vis composition. We carried out low temperature magnetization measurements and the results obtained at 115K are shown in Fig. 4. Further, to verify the quality of our samples and to compare the nature of composition dependence of the magnetic transition temperature with previous reports[7,9], we also carried out temperature dependent magnetization measurements for all compositions. Our results show that the transition temperature varies linearly with composition, and follow the trend reported earlier.[7,9] We also observe that the remanent magnetization of all the phase pure samples increases with increasing Fe content as shown in the inset (a) of Fig 4. This is expected since only Fe and its occupancy on Ga2 site as a consequence of site disordering is responsible for ferrimagnetism in GFO.[9,23] Interestingly, the coercive field ($H_c$), plotted as a function of composition exhibits a maxima at $x = 1.0$ *i.e.* equal Ga and Fe content of the sample, as shown in the inset (b) of Fig. 4. This result is in contrast with that reported by Arima *et al.*[9] on the single crystal samples of GFO who showed a monotonic decrease in the coercivity upon increasing the Fe content of the samples from $x = 0.8$ to $x = 1.1$. While Arima *et al.*[9] attributed this decrease in the coercivity to occupation of Ga sites by Fe ions leading to changes in the magnetic anisotropy, present trend appears to be affected by sample's polycrystallinity. Microcrystallite sizes obtained from XRD data using the Debye Scherrer method (not shown here) suggest a decrease in the crystallite size as the Fe content increases. While the increase in Fe content on one hand tends to increase the magnetic anisotropy as also reported by Arima *et al.*,[9] the decreasing crystallite size and hence decreased size of magnetic domains would impede the magnetic moments and thus leading to the observed behavior. As a further study, it would be interesting to examine the magnetic domains at a microscopic level using techniques such as magnetic force microscopy to further understand the contribution from these two factors.

In conclusion, XRD results clearly showed a phase pure region of $GaFeO_3$ in the range $0.8 \leq x \leq 1.3$ whilst unreacted $Fe_2O_3$ and $Ga_2O_3$ are retained outside this range, clearly marking the phase boundary for pure GFO. Within the boundaries of the phase pure region, the lattice constants vary linearly with the composition with isostructural end members falling on this line suggesting that the system follows Vegard's law excellently. Raman spectroscopy measurements corroborated with the XRD data ruling out any structural change in the samples upon changing the composition. Moreover, shift in the mode frequencies to low frequencies upon increasing Fe content are manifestation of bond stretching, ascribed to a small but finite size difference between Ga and Fe ions. Magnetic measurements revealed that while material's transition temperatures shows a predicted behavior as a function of composition, the coercivity of samples displays a rather different nature than that of single crystals.

*Acknowledgements*

Authors thank Council of Scientific and Industrial Research (CSIR) and Department of Science and Technology (DST), Government of India for the financial support.

**List of Figures:**

Fig. 1 (a) Room temperature X-ray diffraction data with Rietveld refined patterns of $Ga_{2-x}Fe_xO_3$. Pattern for $x = 0.7$ is refined with monoclinic $C2/m$ phase of $\beta$-$Ga_2O_3$ and orthorhombic $Pc2_1n$ phase of $GaFeO_3$ while spectra for $x = 1.4$ is refined using rhombohedral $R\bar{3}c$ phase of $\alpha$-$Fe_2O_3$ and orthorhombic $Pc2_1n$ phase of $GaFeO_3$; (b) spectra of all the compositions within $2\theta$ range $29.5° - 33.7°$ clearly showing the peak shift toward lower $2\theta$ with increasing Fe content (*:$\beta$-$Ga_2O_3$ , \$:$\alpha$-$Fe_2O_3$)

Fig. 2 Variation of lattice parameters of single phase GFO ($0.8 \leq x \leq 1.3$). Parameters of the end members $\varepsilon$-$Fe_2O_3$ (Ref. 16) and $\kappa$-$Ga_2O_3$ (Ref. 15) are represented by diamonds and circles.

Fig. 3 Room temperature Raman spectra of single phase $Ga_{2-x}Fe_xO_3$. For clarity, only 18 modes are marked. Inset shows clearly that the peaks between 600 to 850 $cm^{-1}$ shift towards lower frequency with increasing Fe concentration.

Fig. 4 Magnetization (M) vs. applied field (H) plots for all the single phase compositions of $Ga_{2-x}Fe_xO_3$ measured at 115 K. Inset (a) and (b) show variation of remanent magnetization ($M_R$) and Coercive Field ($H_c$), respectively with Fe content.

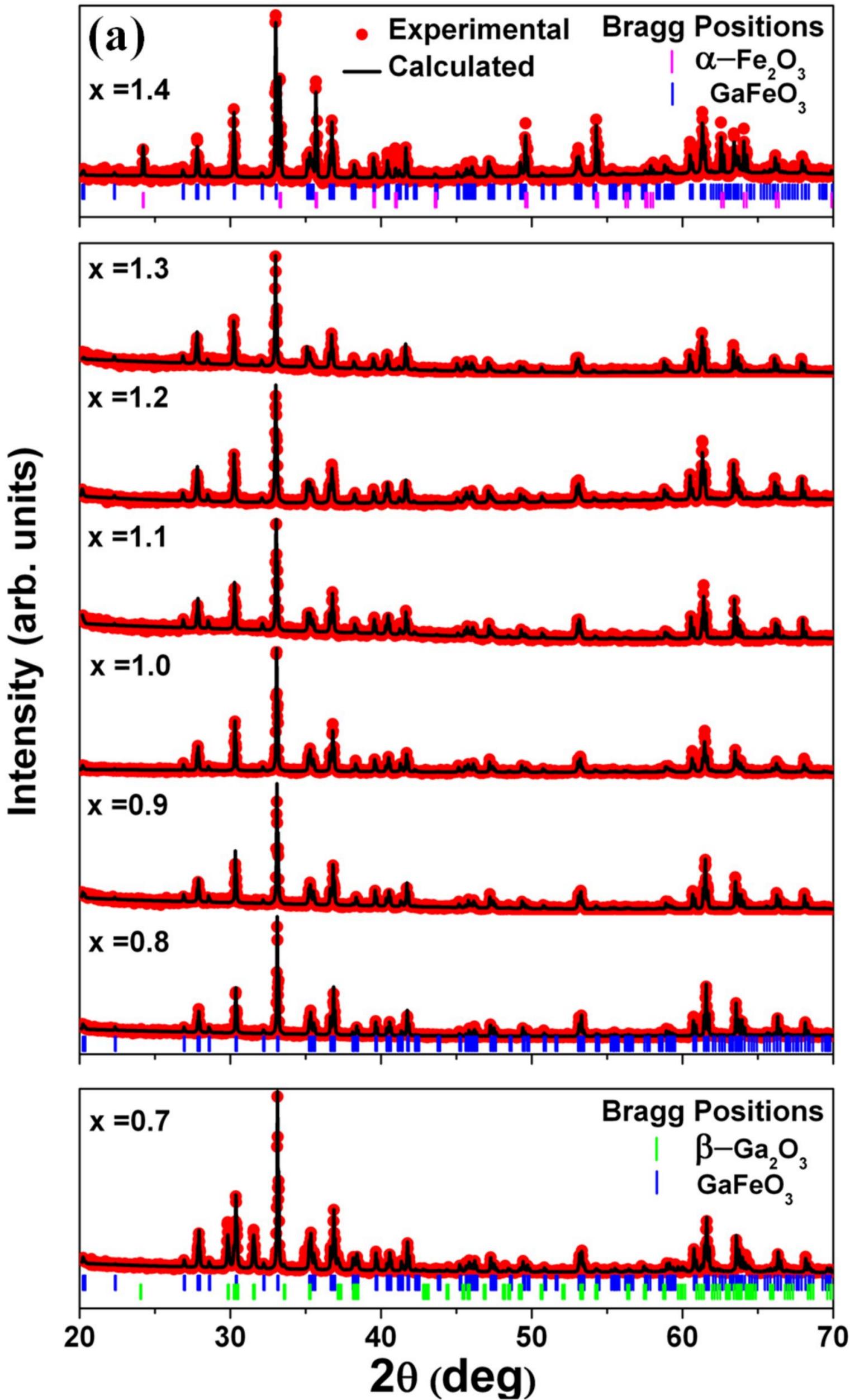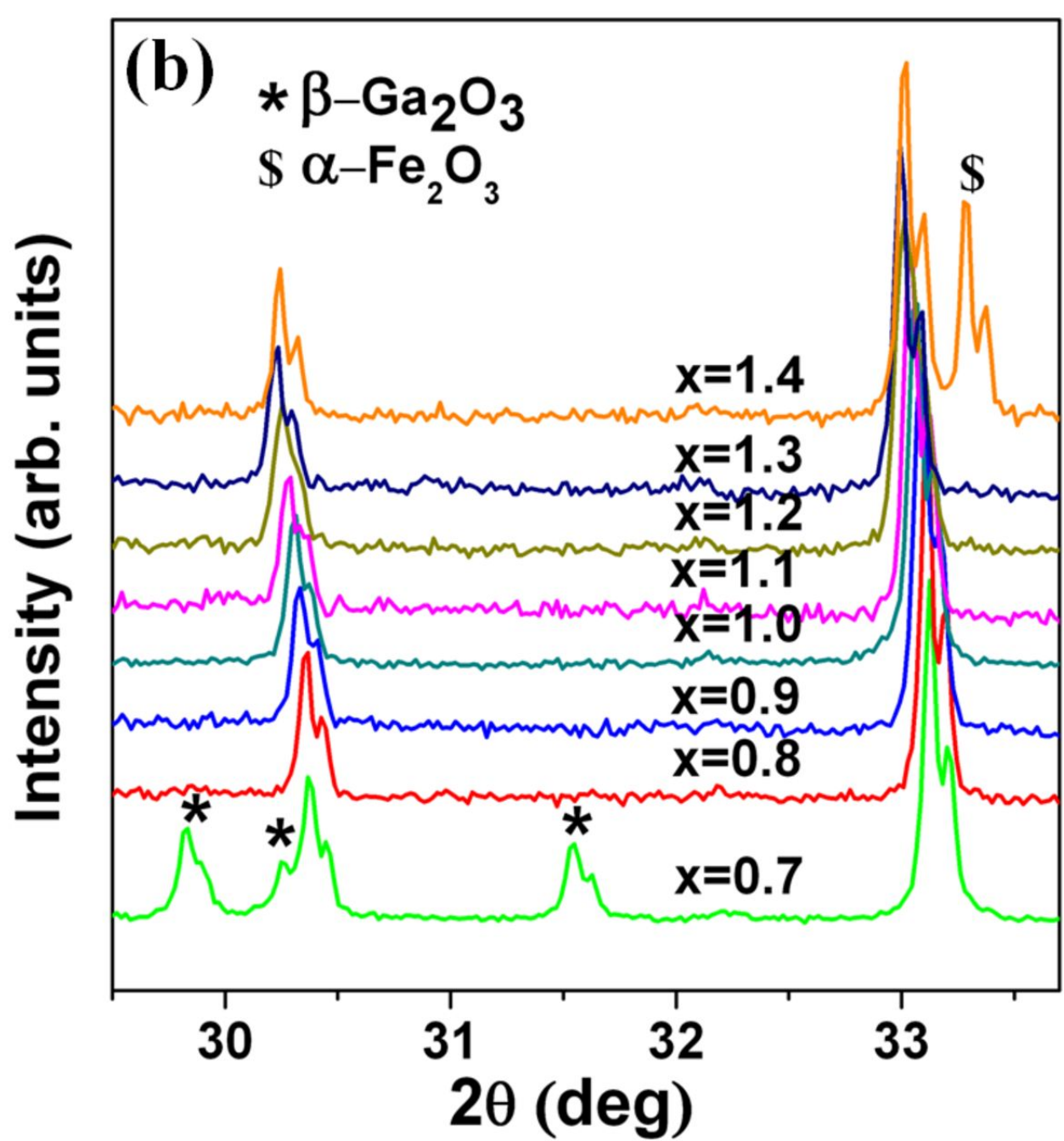

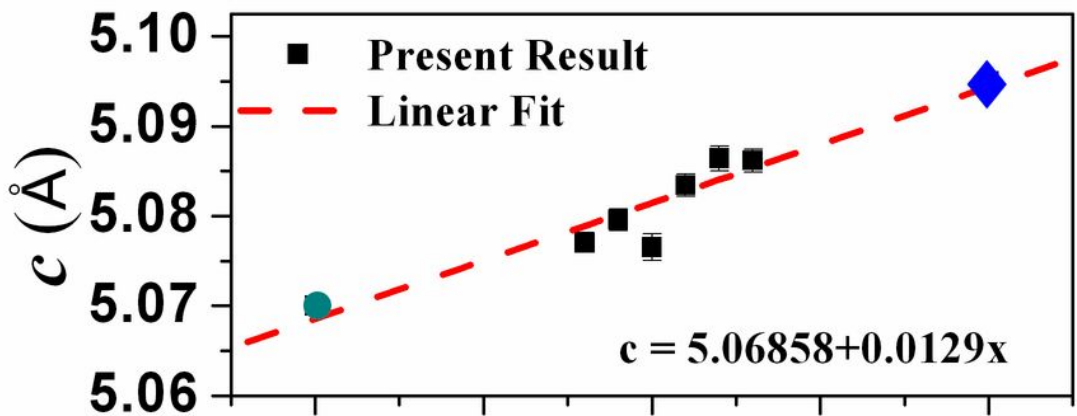
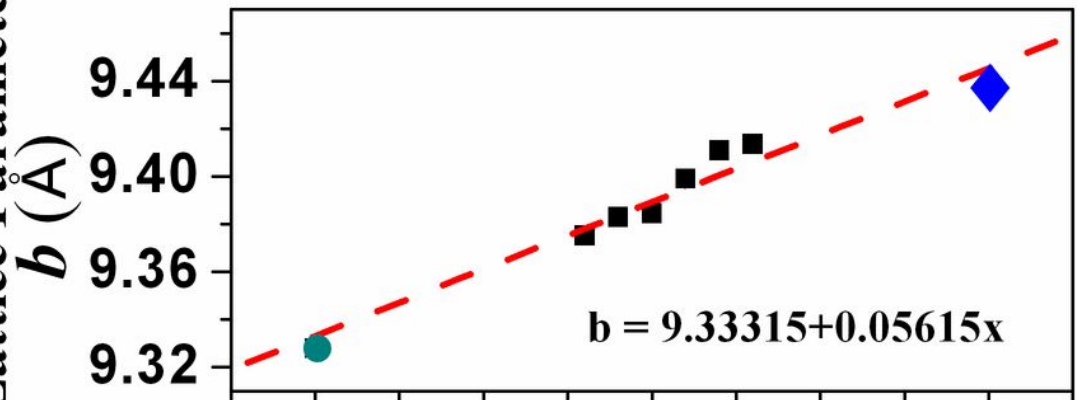
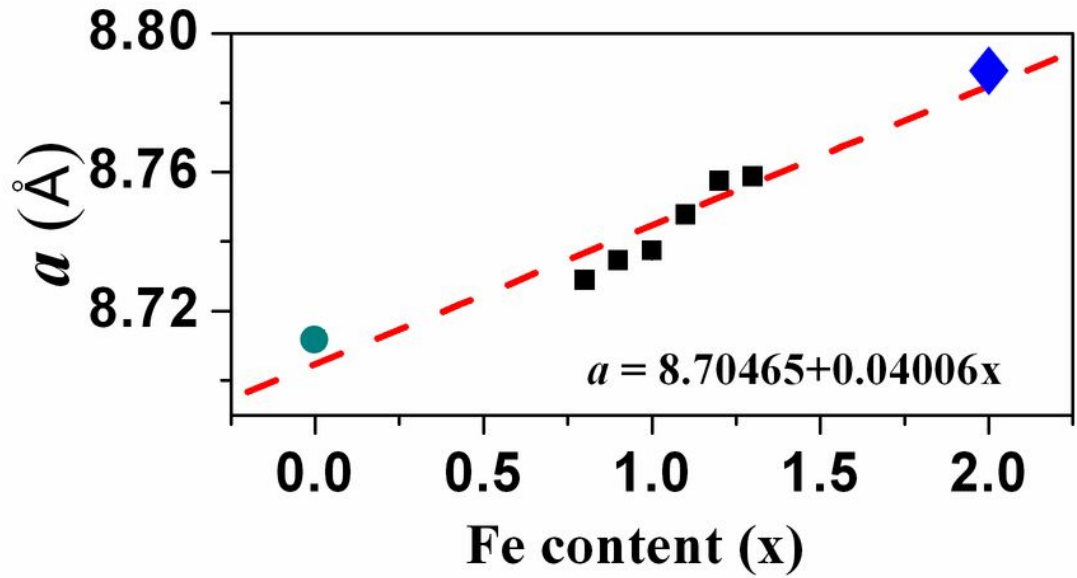

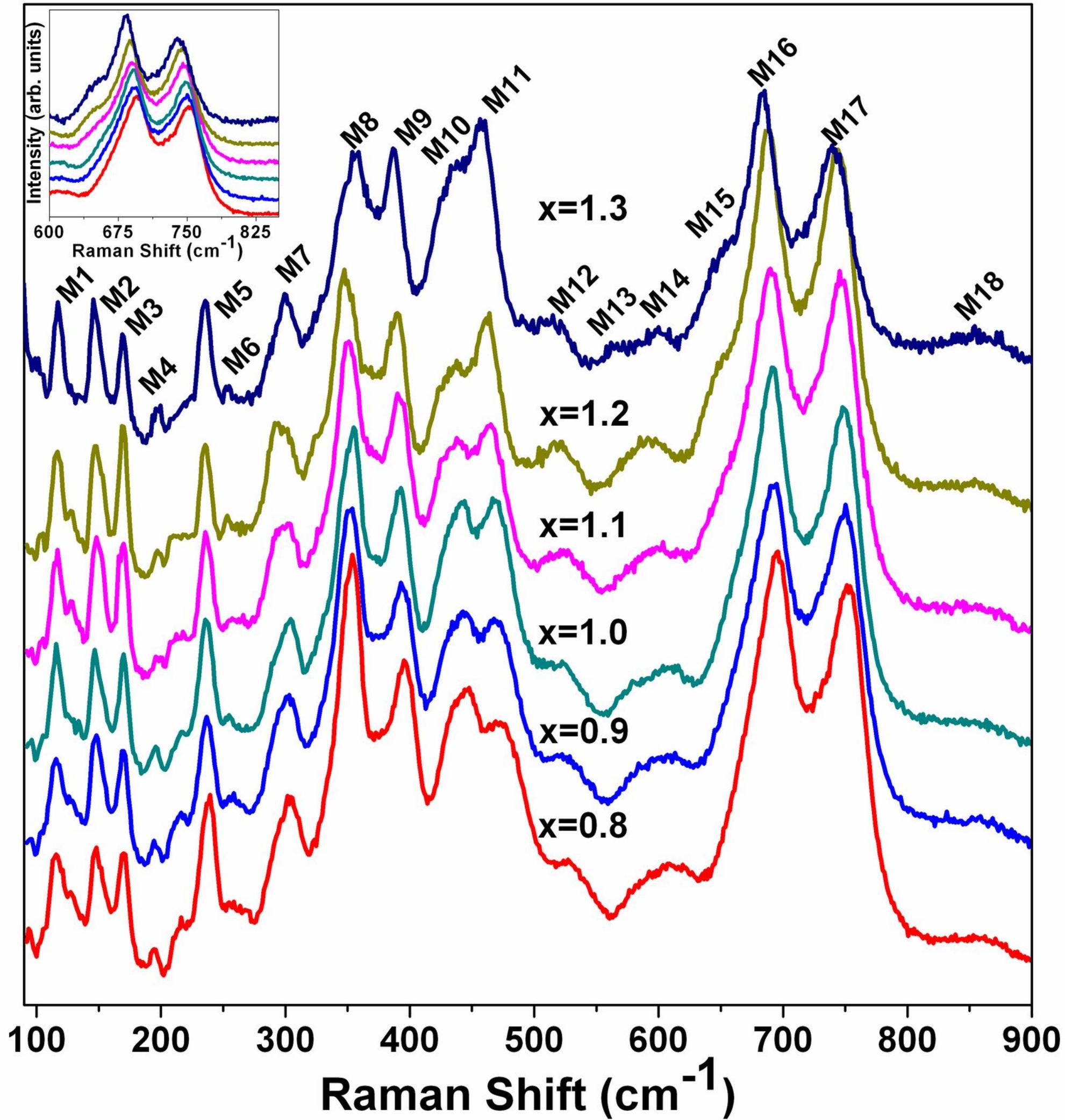

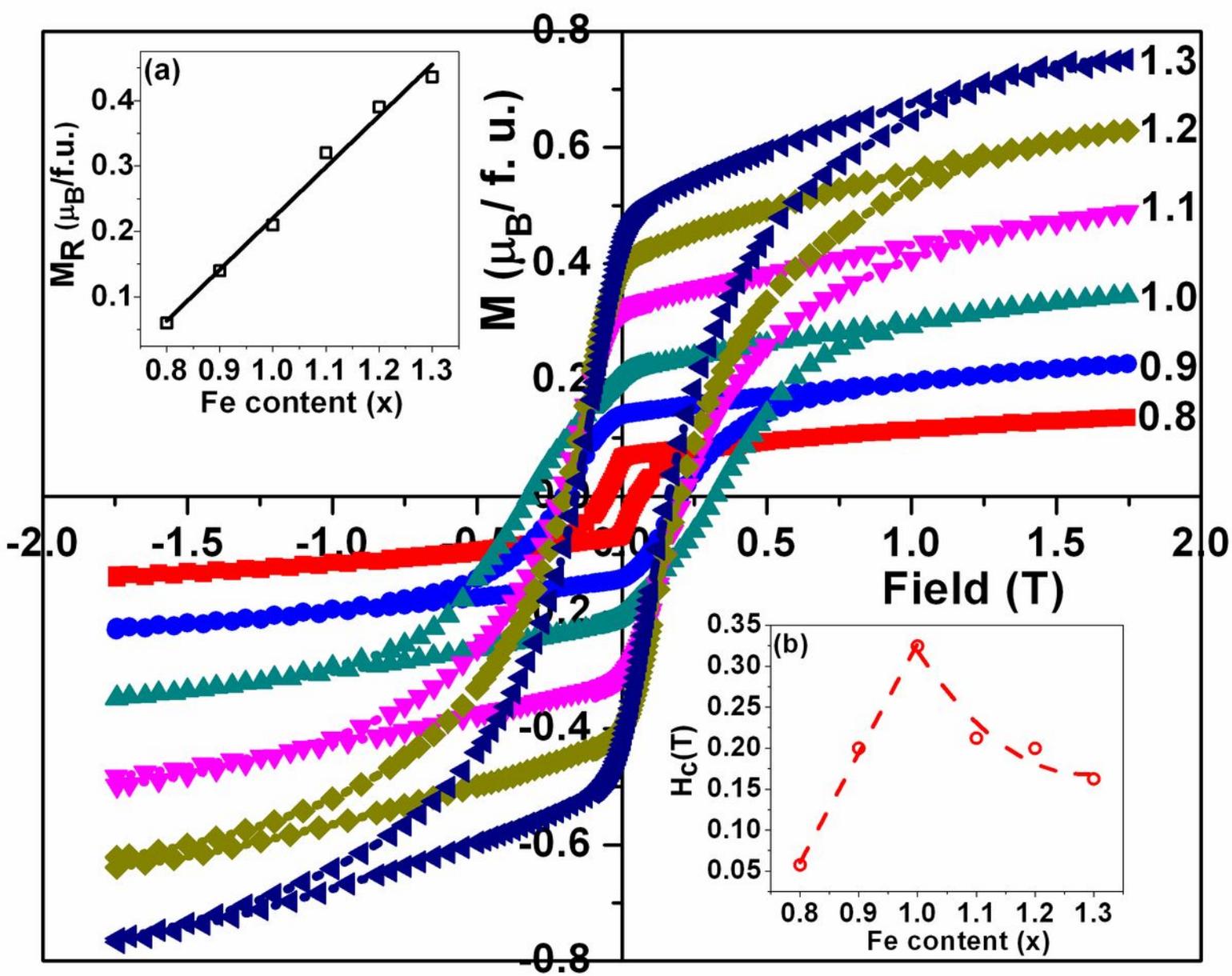